\documentclass[conference]{IEEEtran}
\IEEEoverridecommandlockouts
\usepackage{cite}
\usepackage{amsmath,amssymb,amsfonts}
\usepackage{algorithmic}
\usepackage{graphicx}
\usepackage{textcomp}
\usepackage{xcolor}
\usepackage{hyperref}
\def\BibTeX{{\rm B\kern-.05em{\sc i\kern-.025em b}\kern-.08em
    T\kern-.1667em\lower.7ex\hbox{E}\kern-.125emX}}

\usepackage{caption}
\usepackage{blindtext}
\usepackage{tcolorbox}
\usepackage[final]{pdfpages}
\usepackage{lipsum,multicol}
\usepackage{xcolor}
\usepackage{tikz}
\usepackage{listings}
\usepackage{enumitem}
\usepackage{hyperref}
\usepackage{amsfonts}
\usepackage{wrapfig}
\usepackage{subcaption} 
\usepackage{adjustbox}
\usepackage{colortbl}
\usepackage{fancybox}
\usepackage{multirow}
\usepackage[normalem]{ulem}
\useunder{\uline}{\ul}{}
\usepackage{enumitem}
\usepackage[flushleft]{threeparttable}
\usepackage[T1]{fontenc}
\usepackage{booktabs}
\usepackage{makecell}   % preamble

\newcommand{\ie}{\textit{i.e.,}\xspace}
\newcommand{\eg}{\textit{e.g.,}\xspace}

\newcommand\revision[1]{{\color{blue}{#1}}}

\newtcolorbox{boxK}{
    fontupper = \small,
    sharpish corners, % better drop shadow
    boxrule = 0pt,
    toprule = 0pt, % top rule weight
}

\newcommand{\secref}[1]{Sec.~\ref{#1}\xspace}
\newcommand{\figref}[1]{Fig.~\ref{#1}\xspace}
\newcommand{\tabref}[1]{Table~\ref{#1}\xspace}

\newcommand{\approach}{{\small\texttt{TraceXplainer}}\xspace}

\newcommand{\libest}{\texttt{libEST}\xspace}
\newcommand{\cisco}{\texttt{csc}\xspace}
\newcommand{\albergate}{\texttt{albergate}\xspace}
\newcommand{\ebt}{\texttt{EBT}\xspace}
\newcommand{\etour}{\texttt{etour}\xspace}
\newcommand{\itrust}{\texttt{itrust}\xspace}
\newcommand{\smos}{\texttt{SMOS}\xspace}
\newcommand{\dronology}{\texttt{dronology}\xspace}

\newcommand{\stlibest}{ST1\xspace}
\newcommand{\stcisco}{ST2\xspace}
\newcommand{\stalbergate}{ST3\xspace}
\newcommand{\stebt}{ST4\xspace}
\newcommand{\stetour}{ST5\xspace}
\newcommand{\stitrust}{ST6\xspace}
\newcommand{\stsmos}{ST7\xspace}
\newcommand{\stdronology}{ST8\xspace}

\newcommand{\exbase}{$EX_0$\xspace}
\newcommand{\exfirst}{$EX_1$\xspace}
\newcommand{\exsecond}{$EX_2$\xspace}
\newcommand{\exthird}{$EX_3$\xspace}
\newcommand{\exfourth}{$EX_4$\xspace}

\newcommand{\scm}{SCM\xspace}
\newcommand{\wmd}{WMD\xspace}
\newcommand{\cosine}{COS\xspace}
\newcommand{\euc}{EUC\xspace}

\newcommand{\wv}{\textit{word2vec}\xspace}
\newcommand{\dv}{\textit{doc2vec}\xspace}

\newcommand{\bits}{\textit{$\mathcal{B}$}\xspace}

\usepackage{xcolor}

\definecolor{codegreen}{rgb}{0,0.6,0}
\definecolor{codegray}{rgb}{0.5,0.5,0.5}
\definecolor{codepurple}{rgb}{0.58,0,0.82}
\definecolor{backcolour}{rgb}{0.95,0.95,0.92}

\lstdefinestyle{mystyle}{
  backgroundcolor=\color{backcolour}, commentstyle=\color{codegreen},
  keywordstyle=\color{magenta},
  numberstyle=\tiny\color{codegray},
  stringstyle=\color{codepurple},
  basicstyle=\ttfamily\tiny,
  breakatwhitespace=false,         
  breaklines=true,                 
  captionpos=b,                    
  keepspaces=true,                 
  numbers=left,                    
  numbersep=3pt,                  
  showspaces=false,                
  showstringspaces=false,
  showtabs=false,                  
  tabsize=2
}

\renewcommand{\arraystretch}{1.2}

\begin{document}

\title{
Lost in Transmission: An Information-Theoretic Account of Unsupervised Software Traceability
%On Interpreting the Effectiveness of Unsupervised Software Traceability with Information Theory
%Using Information Theory to Interpret Unsupervised Software Traceability
}

\author{\IEEEauthorblockN{Daniel Rodriguez-Cardenas\textsuperscript{*}, Logan Fecko, Denys Poshyvanyk   }
\IEEEauthorblockA{
\textit{William \& Mary}\\ Williamsburg, VA, USA\\
\{dhrodriguezcar, lefecko, dposhyvanyk\}@wm.edu}
\and
\IEEEauthorblockN{  David N. Palacio\textsuperscript{*} }
\IEEEauthorblockA{%\\
\textit{Microsoft}\\ Redmond, WA, USA \\
davidnad@microsoft.com}
\and
\IEEEauthorblockN{Kevin Moran}
\IEEEauthorblockA{%\\
\textit{University of Central Florida}\\ Orlando, FL, USA \\
kpmoran@ucf.edu}

}

\maketitle

\begin{abstract}
Traceability remains a critical capability to ensure system reliability, maintainability, and compliance in modern software development. Although unsupervised Information Retrieval (IR) and Machine Learning (ML) techniques are widely adopted for automated trace link recovery, their effectiveness is often limited by the quality and structure of the underlying artifacts. In practice, these approaches assume that meaningful traceability signals are embedded in textual data, an assumption that rarely holds in industrial settings with sparse, inconsistent, or unbalanced documentation. Furthermore, conventional evaluation metrics (\eg precision, recall, F1) can misrepresent performance when data characteristics are not explicitly considered. We introduce \approach, an information-theoretic framework for evaluating the reliability and limits of unsupervised traceability. Our approach leverages \textit{self-information} and \textit{mutual information (MI)} to quantify the informativeness and alignment of source and target artifacts. Through a comprehensive empirical analysis of industry datasets, we show that typical traceability corpora exhibit significant information imbalances, where the source code contains on average more information than the corresponding documentation. In addition, the observed levels of mutual information, loss, and noise reveal inherent constraints on the ability of unsupervised techniques to recover accurate trace links. These findings suggest that improving traceability in practice requires a shift to data-centric engineering, focusing on artifact quality, consistency, and information alignment; rather than solely advancing model sophistication (or complexity). Our results provide insights for practitioners to better assess traceability readiness and guide improvements in documentation and development workflows.

\end{abstract}
\thanks{\footnotesize\textsuperscript{*}Authors contributed equally.}


\section{Introduction}\label{sec:introduction}

In this paper, we examine the phenomenon of information transmission in software traceability from the perspective of industrial software engineering practice. Traceability, the discipline of drawing semantic relationships among software artifacts (\eg code, requirements, test cases), is a foundational capability in modern engineering organizations. These semantic relationships underpin everyday practitioner workflows including code comprehension~\cite{moran_improving_2020}, compliance validation, security tracking~\cite{palacio_security,Gadelha2021TraceabilityRB}, and impact analysis~\cite{Aung2020ALR,Falessi20}. In industrial settings, information retrieval techniques (IR) (\eg TF-IDF, LSA, or LDA) are routinely employed to mathematically represent \textit{high-level} artifacts (\ie requirements) and \textit{low-level} artifacts (\ie source code) in compressed tensors~\cite{Dit2013,Ge2012,Dasgupta2013,Dit2013a}. These tensors are generated via unsupervised learning and used to compute a distance (\eg Euclidean, Cosine, or Word Mover's Distance) between two artifacts in a derived vector space. The distance defines how semantically close two artifacts are to each other, and unsupervised traceability pipelines exploit these distances to confirm whether a \textit{source-target} data sequence pair should be linked in a production trace matrix.

%%% Background
In industry, the effectiveness of unsupervised software traceability is typically reported in terms of canonical classification metrics such as precision, recall, AUC, accuracy, or F1. However, these metrics can be misleading when the underlying data are not properly explored and analyzed, a recurring concern for engineering teams that must justify automated trace links to auditors, release managers, and compliance owners. In real software traceability corpora, datasets are generally imbalanced, skewed, and biased, as we demonstrate in our empirical study. This observation is consistent with previous calls for more rigorous statistical analysis in software engineering research~\cite{moran_improving_2020,Kitchenham2017RobustEngineering}. Consequently, we contend that there are \textit{data limitations} embedded in the software artifacts that traceability techniques operate upon and that these limitations cap the effectiveness of any tool deployed on top of them.

This paper develops techniques that automatically articulate the cases in which the information captured by industrial datasets causes unsupervised methods to trace links ineffectively, resulting in erroneous predictions that propagate into downstream engineering activities. Consider, for example, a development team that must assess the impact of incoming new requirements ahead of a release. How should practitioners proceed if they are unfamiliar with the code-based components of the system under analysis? They will most likely begin by reading the associated documentation (\ie code comments) to \textit{trace} the functionality between requirements and code. Yet \textit{what happens if requirements are poorly written? Or if certain areas of the source code are not completely documented or are obfuscated?} The traceability process --and, by extension, the impact-assessment task it supports-- becomes cumbersome, inefficient, and a source of operational risk. Our aim is to use information science together with automated traceability to identify the artifacts that most degrade system understandability in such industrial workflows.

%% Stating the Problem
Treating \textit{data} as the central aspect of learning theory, insufficient or poorly curated data is a deficiency that is directly reflected in the effectiveness of unsupervised algorithms~\cite{bishop_deep_2024}. Unfortunately, neither IR techniques nor, more generally, unsupervised models can guarantee that a \textit{predicted} trace link is reliable enough to be acted on in production. Because there is a data-dependent gap among software artifacts, unsupervised traceability remains a difficult and error-prone task in industrial deployments. Consequently, \textit{bridging the traceability research gap is infeasible under the assumption that the data is poorly structured or of low quality}. Traceability practice therefore requires efficient statistical approaches that quantify how \textit{reliable} predicted trace links actually are before they are surfaced to engineers, auditors, or release tooling.

%% Hypothesis
Textual artifacts produced in real engineering environments may not contain enough information for an unsupervised technique to learn patterns that determine a trace link. We hypothesize that information-theoretic measures (\eg self-information, mutual information, relative entropy, and shared information) can help \textbf{interpret or explain} how unsupervised techniques are \textit{limited for solving the traceability problem} in industrial settings. Unsupervised models (\ie conventional, machine learning, or even Large Language Models) are limited by the informativeness of the data they operate on. We evaluate this limitation empirically only for classical embedding-based unsupervised techniques (\wv, \dv); whether supervised, transformer-based, or LLM-based methods can narrow this gap using contextual or world knowledge remains an open question (see \secref{sec:lessons}). By performing an information-theoretic analysis, we can obtain a diagnostic estimate of how robust, reliable, and trustworthy unsupervised techniques are likely to be under real-world data conditions, offering practitioners a risk indicator rather than a definitive verdict on any individual technique or link. We named this analysis \approach, an interpretability approach that gives a starting point for monitoring traceability distances, information-theoretic measures, and the relationship between distances and information measures.

%% The goal of the study
This research aims to demonstrate that traceability models are limited when they rely only on unsupervised learning and textual data, and that these limitations are best understood and addressed from a data-centric, industry-grounded perspective. Through an extensive (information-based) empirical analysis, we show how unsupervised techniques are inadequate for several training configurations and data modalities, due to ineffective predictions on imbalanced or under-informative artifacts. Our work is therefore a data-centric analysis aimed at characterizing the quality of the testbeds employed and reported in the traceability literature, with particular attention to those drawn from industrial pipelines. A brief overview of our findings from the exploratory analysis is provided below.

For the \cisco system, an industrial testbed derived from a real Cisco engineering workflow, we found that the pull request comments and the associated source code often contain contrasting information. The pull request comments exhibit an entropy of $3.42[0.02]\,\bits$, while the source code comments reach $5.91[0.01]\,\bits$, indicating that the two sides of a change describe substantially disparate information. The average level of mutual information was $3.21\,\bits$ (error $0.02$ at a $95\%$ confidence level). This imbalance suggests that code changes do not accurately capture the information content of the originating pull request at implementation time; a pattern that has direct implications for impact analysis, code review, and downstream audit activities. In practice, we recommend that engineering teams examine such imbalanced links, and the information discrepancies between pull requests and source code, in order to design refactoring strategies that reduce information loss and increase mutual information across the software documentation. Our complete empirical study contributes industry-oriented results on unsupervised models, a complete analysis of information transmission across testbeds, and a correlation study between semantic distances and information metrics, all framed to be actionable for engineering teams adopting (or already operating) automated traceability solutions.

\section{Background \& Related Work}\label{sec:background}

Software requirements should be capable of being \textit{translated} into multiple forms of information, such as source code, test cases, or design artifacts. Thus, we refer to these requirements or any initial/raw form of information as the \textit{source} artifacts. In contrast, the information product of a transformation or alteration is considered a \textit{target} artifact. In the software engineering context, a transformation could be any action that software engineers or practitioners apply from those requirements to source code. For example, implementing a requirement can be seen as a way to \textit{translate} information from requirements to software components. Our research deals with the intersection of software traceability and interpretability techniques to introduce statistical and information theory methods to explain the ineffectiveness of unsupervised traceability. 

\textbf{Software Traceability Techniques.} Traceability datasets consist of corpora of textual artifacts and source code files, where the goal is to establish links between source documents and target code elements. Early approaches primarily relied on \textit{Information Retrieval} (IR) techniques (\eg Vector Space Models, Jensen–Shannon divergence, and Topic Models) to recover such links~\cite{dit,Dit2013,Dit2013a,Dit2013b}. Stronger, paradigms include supervised deep-learning rankers~\cite{Guo2017SemanticallyTechniques}, pre-trained Transformer and code-language-model rerankers~\cite{Lin2021TraceabilityTransformed,Zhang2023EALink,UniXcoder}, graph-based methods that exploit structural or dependency relations~\cite{Zou2024HANTracer}, and, most recently, LLM/RAG-based retrieval-augmented approaches~\cite{Fuchss26}, in addition to earlier representation-learning work~\cite{Zhao2017WordEmbeddingsTraceability}; these explicitly learn semantic similarity and structural patterns from labeled links, often yielding substantial improvements in requirements-to-code and issue–commit traceability tasks. However, most existing techniques implicitly assume that the distributional and structural properties of tokens (\eg words or identifiers) are comparable across requirements and source code. This assumption is problematic: although source code can be represented as a sequence, it is significantly more constrained, repetitive, and structurally governed than natural language~\cite{Karampatsis2020BigC,Devanbu12}. Consequently, many unsupervised methods (\eg \wv, \dv, autoencoders, and Transformers)~\cite{palacio_security,Lin2022EnhancingAS,Du2020AutomaticTL} may capture spurious or unstable similarities, which is reflected in their high performance variability across datasets. Despite these limitations, little work has explicitly examined the role of entropy or other information-theoretic measures in exploratory traceability analysis. 

Prior research has largely overlooked such metrics when evaluating unsupervised approaches, although some studies emphasize the importance of modeling software artifacts as probability distributions~\cite{moran_improving_2020}. 

T-BERT~\cite{Lin2021TraceabilityTransformed} supports  our thesis as evidence that external data can offset local information limits, consistent with our data-centric perspective. Much of the gain comes from pretraining on CodeSearchNet’s function–docstring pairs and lies outside the artifact-level signal our approach measures. T-BERT evaluates ranking over small candidate sets ($\sim$50 artifacts), unlike \approach, which uses a large, imbalanced classification setting (\eg $21$K pairs for \cisco), so the results are not directly comparable.

\textbf{Interpretability Techniques.} Interpretability has been used to explain the predictions generated by (deep) learning models, complementing traditional evaluation methods, and enhancing our understanding of the decision-making process to reduce uncertainty. Post-hoc interpretability assumes that the model is a black box and analyzes its behavior by examining changes in the output derived from the input. In  Machine Learning, most research has focused on improving the plausibility and faithfulness of explanations such as LIME \cite{ribeiro_why_2016}, DeepLIFT \cite{shrikumar_learning_2019}, Shapley values \cite{lundberg_unified_nodate}, and more recent model-specific based approaches tailored for large language models \cite{UnlockingBlackBox}. Information theory configures the interpretability method to better understand the model's decision when the model traces the source and target artifacts. To the best of our knowledge, our work is the first to evaluate unsupervised techniques for the traceability problem with information theory.

\section{Information Theory for Traceability}\label{sec:approach}

\begin{figure}[h]
		\centering
    \includegraphics[width=0.45\textwidth]{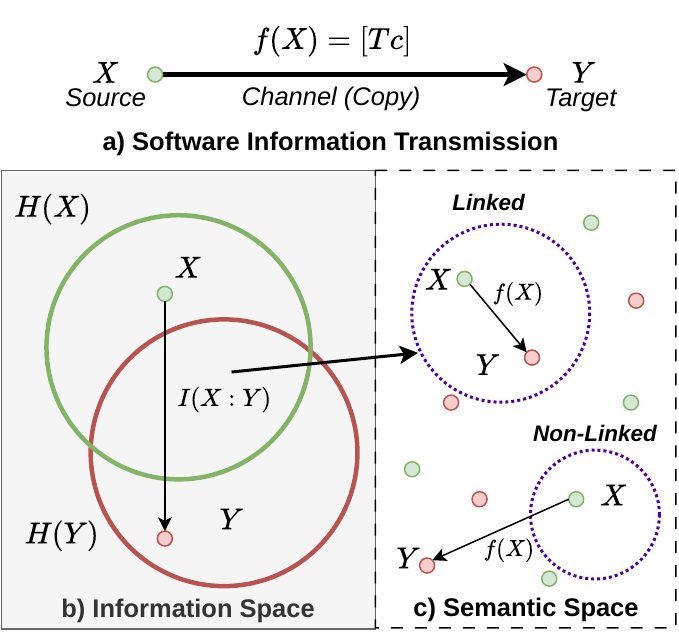}
		\caption{\approach: Using Information Theory to Interpret Unsupervised Traceability Models.}
        \vspace{-0.3cm}
        \label{fig:tracexplainer}
\end{figure}

%%Introducing the solution
This section introduces the information-theoretic foundations underlying our approach \approach, \a data-centric diagnostic lens for software traceability. \approach computes a set of information measures that complement and explain the limitations of semantic traceability techniques, enabling us to assess how well traceability algorithms perform on arbitrary software projects. Our experiments show that studying the \textit{manifold} of information measures (\ie the information and semantic spaces) helps surface critical points in the artifacts: undetected links typically appear in regions where the information content of two artifacts $h(x)$ and $h(y)$ overlaps (see \figref{fig:info-theory}), often due to incomplete documentation or redundant tokens; both candidates for refactoring. \figref{fig:tracexplainer} depicts the three components of the approach: software information transmission, the information space, and the semantic space.

%% Entropy Explanation
\subsection{Software Information Transmission}

Probability theory and \textit{information theory} provide the foundation on which we build unsupervised interpretability for software traceability. Information theory, in particular, lets us quantify the amount of information present in software corpora. Intuitively, the information content of a software artifact reflects its \textit{degree of novelty}: observing the value of a discrete random variable $x$ is more informative when $x$ is improbable than when it is expected. The content therefore depends on the underlying probability distribution $p(x)$ through a monotonic function $h(x) = -\log_2 p(x)$. Higher $p(x)$ (\ie more frequent values) yields lower information. Units of $h(x)$ are \textit{bits} (\ie binary digits $\bits$) when the logarithm is taken in base two, and \textit{nats}~\cite{bishop_deep_2024} when expressed in natural logarithms.

Software artifacts (\eg code, requirements, tests) are sequence-based structures that record the outputs of distinct generation processes such as programming, requirements elicitation, or issue management. By observing specific values of $x$, we can monitor how much information is transmitted. A first-order representation treats an artifact as the relative frequency of its tokens $t$: $p(x = t_i) = t_i/|t|$, where $t_i$ is the count of token $i$ in vocabulary $\mathbf{v}$ and $|t|$ is the sequence length. Letting a source (\ie sender) transmit artifact $x$ to a target (\ie receiver), the expected information content is the \textit{entropy} of $x$:
$H[x] = -\sum_x p(x) \log_2 p(x)$. A software engineering process can thus be characterized as producing artifacts carrying $m$ bits of sequence-based information.

Information can flow from source to target by \textit{copy} $[T_c]$ or by \textit{transformation} $[T_s]$, and the total transmission is captured by the \textit{mutual information} $I(x{:}y)$. Critically, $I(\cdot)$ alone cannot tell these two regimes apart. Because the atomic unit of a software artifact is its set of tokens, transmission is by copy when the target $y$ preserves the source token set, and by transformation when $y$ undergoes a systematic syntactic or semantic modification of $x$~\cite{Kolchinsky2020DecomposingTransformation}. Disentangling the transformation regime, for instance via the decomposition\cite{Kolchinsky2020DecomposingTransformation}, is an open problem at the intersection of software engineering and information retrieval; this work contributes to bridging the gap between informativeness and unsupervised traceability learning.

Source artifacts express \textit{intent}, while targets encode \textit{implementation} via APIs, control flow, and abstractions that rarely reuse the source vocabulary; for instance, a requirement mentioning \textit{loop} may be implemented with \texttt{for}, yielding near-zero token overlap despite a valid link. Because $I(x{:}y)$ operates on token distributions, it can \textbf{underestimate} such semantically valid but lexically divergent links, biasing loss and noise toward labeling well-transformed artifacts as under-documented. We do not model this transformation explicitly; accordingly, our measures characterize \textbf{lexical (token-level) traceability}. A target’s vocabulary is explained by its source rather than by full semantic or behavioral traceability.

\subsection{Software Information Space}
\begin{wrapfigure}{r}{0.28\textwidth}
%\begin{figure}[hb]
 \vspace{-0.8em}
		\centering
    \includegraphics[width=0.28\textwidth]{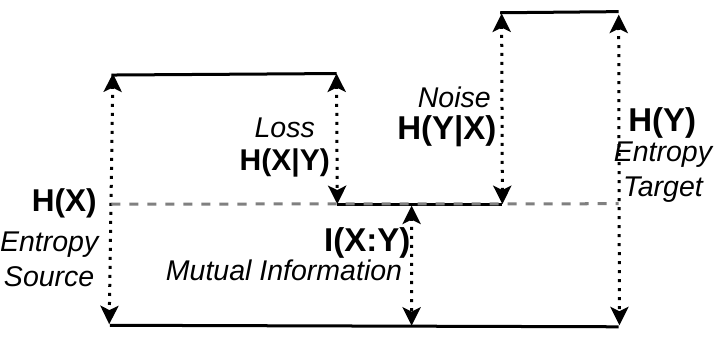}
		\caption{Information Theory Measures in \approach.}
        %\vspace{-0.3cm}
        \label{fig:info-theory}
        \vspace{-0.8em}
%\end{figure}
\end{wrapfigure}

This subsection focuses on the \textit{information manifold}, the lens through which \approach interprets traceability. We deliberately separate this view from the \textit{semantic space}; the formal mathematical specification of how unsupervised techniques (\eg \wv, \dv, soft cosine, or word mover's distance) detect candidate links, which we already covered in the background literature~\secref{sec:background}. Whereas the semantic space explains \textit{how} similarity is computed, the information space explains \textit{how much information is transmitted} between source and target artifacts, and \textit{where} that transmission breaks down. \figref{fig:info-theory} depicts the interactions of \textit{entropy} between two artifacts (\ie source and target). These interactions give rise to related informativeness measures: self-information, information loss, information noise, and \textit{mutual information}~\cite{MacKay2003InformationAlgorithms}, which we describe below.

\textbf{Self-Information of Source Artifacts $[H(X)]$} quantifies the amount of encoded information (\ie in $\bits$ bits) carried by the sender's artifacts, such as requirements, issues, or pull requests. Throughout this paper, ``self-information of an artifact'' denotes the expected self-information of its token distribution, \ie its Shannon entropy $H(X)$, as opposed to the pointwise self-information $h(x)$ of a single token defined in \secref{sec:approach}. Consider, for instance, the industrial dataset \cisco, which contains a pull request whose body is reduced to just two tokens: \texttt{"B dtimeout"}. The self-information of such a sparse artifact collapses to $H(x)=0.0\,\mathcal{B}$, indicating no recoverable information. In general, source artifacts composed of few or highly repetitive tokens depress entropy estimates. Tracking the sender's information is therefore a prerequisite for explaining \textit{how traceability occurs}.

\textbf{Self-Information of Target Artifacts} $[H(Y)]$ intuitively captures the amount of information contained in the receiver's artifacts, such as source code, test cases, or configuration files. The snippet below shows a standard exception drawn from \cisco, whose self-information value is $H(y) = 4.16\,\mathcal{B}$. Note that the previously discussed source content is less diverse than this target, assuming the two artifacts are linked. This measure is most useful when compared against the corresponding source, since the comparison detects imbalanced trace-link information.

\begin{center}
\begin{minipage}{0.47\textwidth}
\begin{lstlisting}[language=Python, basicstyle=\footnotesize\ttfamily, aboveskip=2pt, belowskip=2pt, xleftmargin=1em]
import sys
import traceback
def fireException(message):
    try:
        with open("buddy_script_error.txt", "w") as file:
            file.write(str(message))
        print(message)
        traceback.print_exc()
        sys.exit(-1)
    except IOError:
        traceback.print_exc()
        sys.exit(-1)
\end{lstlisting}
\end{minipage}
\end{center}

\textbf{Mutual Information $[I(X{:}Y)]$} or \textit{MI}
embodies the amount of information that source $X$ shares with target $Y$ (and, complementarily, that $Y$ shares with $X$). \textit{MI} quantifies the information artifacts hold in common (\ie the information transmitted from a source to a target). For example, the \cisco's trace link $\{PR_{256} \to binaryfunc.py\}$ (\ie pull request to code) has an \textit{MI} of $6.38\mathcal{B}$, suggesting the source artifact's content overlaps a large portion of the target's. In our \textit{Case Study \revision{3}} (see \secref{sec:cases}), we demonstrate that the potential link
$\{PR_{56} \to binaryfunc.py\}$ registers the same amount of MI as the previous example, but it was omitted in the ground truth. Hence, \textit{MI} helps practitioners verify trace links for benchmarking purposes.

\textbf{Information Loss} $[H(X|Y)]$ quantifies the amount of information that enters the \textit{channel} but does not leave it; that is, information in the source that cannot be recovered from the target. In a software engineering context, this typically arises from either (1) missing code implementation or (2) poorly documented code or requirements. For example, the links $\{PR_{240} \to \texttt{binaryfunc.py}\}$, $\{PR_{240} \to \texttt{securityfunc.py}\}$, and $\{PR_{168} \to \texttt{auth.py}\}$ exhibit losses of $6.51\,\mathcal{B}$, $6.48\,\mathcal{B}$, and $6.3\,\mathcal{B}$, respectively. These \cisco pull requests share a low self-information value, indicating that they were poorly described in natural language~\cite{palacio2023tracexplainer}.

\textbf{Information Noise} $[H(Y|X)]$ quantifies the amount of information that leaves the channel without having entered it; that is, information in the target that cannot be traced back to the source. In practice, this corresponds to implemented code that is not described in the originating pull request or requirements. For example, the links $\{PR_{194} \to \texttt{\_\_init\_\_.py}\}$, $\{PR_{177} \to \texttt{fireException.py}\}$, and $\{PR_{56} \to \texttt{setup.py}\}$ exhibit noise of $1.83\,\mathcal{B}$, $1.81\,\mathcal{B}$, and $1.55\,\mathcal{B}$, respectively. These Python files present low entropy values because they are configuration scripts, suggesting that the corresponding ground-truth link associations are either incorrect or incomplete~\cite{palacio2023tracexplainer}.

\textbf{Minimum Shared Information} (\textit{MSI}), reported in this work both as entropy $Si(X{:}Y)$ and as extropy $Sx(X{:}Y)$ (\ie measures the missing information in a given software artifact). \textit{MSI}  differs from \textit{MI} in that it considers only the minimum token overlap between two artifacts when estimating the entropy. For instance, consider an artifact $A$ with token counts $A = [(\text{for}, 14), (\text{if}, 3), (\text{return}, 10)]$ and an artifact $B$ with token counts $B = [(\text{for}, 10), (\text{if}, 0), (\text{return}, 20)]$. The minimum shared vector is $\min(A, B) = AB = [(\text{for}, 10), (\text{if}, 0), (\text{return}, 10)]$. This shared vector defines a probability distribution from which the minimum entropy and extropy can be computed, helping to detect artifacts that share no tokens (\ie null minimum shared information). For example, \cisco contains $5{,}417$ potential null-shared links, of which only $68$ match the ground truth.

%\begin{itemize}
%\item Examples of PR-295: 
%${PR-295 ➝ psb_mapping.py}, {PR-285 ➝ binaryScan.py}, {PR-285 ➝ test_BinaryScan.py}

%\item Examples of PR-241: 
%${PR-241 ➝ csbcicd_report/csbcicd_func.py}, {PR-241 ➝ third_party/binary_scan_func.py}, {PR-241 ➝ csbcicdReport/test_csbcicd_func.py}
%\end{itemize}

%MISCELLANEUS <<------ @danaderp

%We assume $p(x)$ distributions are not influenced by any type of prioritization. There is no preference for selecting the token \texttt{public} over the token \texttt{private} if our experiment is to try to generate a token in a sequence. Or we can see probabilities as frequencies. So we can relate all the possible token combinations to a pattern. This pattern can be measured. 

%Finally, to calculate the self-information of some document X, we compute:

%$H(x)= - \sum_i P_x(x_i)log_2 P_x(x_i) $

%where $P_x(x_i)$ is the probability of some token $x_i$ appearing in document $x$

%For our purposes, a corpus of objects can have different granularity. The lowest granularity would be a corpus of words or tokens inside a source or target document, e.g., a Java source code file. The next granularity up would be a corpus of multiple source or target documents of the same artifact type, i.e., all high-level requirement documents. The last granularity involves a corpus of both sources and targets. Each granularity allows for different questions to be asked. 

\section{Empirical Design}\label{sec:design}

Our study focuses on leveraging \approach to interpret the (negative) effectiveness of traceability techniques. To this end, we define a set of \textbf{RQs} that investigate informativeness measures across software artifacts and examine how these measures correlate with traceability effectiveness, with the goal of \textit{interpreting unsupervised techniques}.

\begin{enumerate}[label=\textbf{[RQ$_{\arabic*}]$}, ref=\textbf{RQ$_{\arabic*}$}, wide, labelindent=5pt]\setlength{\itemsep}{0.2em}
      \item \label{rq:effectivess} \textit{How effective are unsupervised techniques at predicting candidate trace links using IR/ML representations?}
      \item \label{rq:semantic} \textit{To what extent are semantic metrics imbalanced to the ground truth?}      
      \item \label{rq:exploratory} \textit{How much information is transmitted from source to target artifacts?}
      \item \label{rq:correlation} \textit{To what extent do information metrics correlate with semantic distances?}
\end{enumerate}

\subsection{Experimental Context}

To answer research questions, we designed an experimental framework comprising a collection of datasets, trained models, and testbeds to evaluate \approach.

\textit{Datasets and Training Models.} To evaluate our metrics for traceability recovery, we pretrained our models while varying three factors: the preprocessing strategy, the vectorization technique, and the pretraining dataset. The preprocessing strategy was one of a conventional unsupervised tokenizer based on NLTK~\cite{bird-loper-2004-nltk}, bpe8k, or bpe32k, where bpe8k and bpe32k are subword units produced by the SentencePiece tokenizer~\cite{sentencepiece}. For vectorization, we used the skip-gram for \wv and the bag-of-words (pv-bow) vector of the paragraphs for \dv. Pretraining was conducted on CodeSearchNet for Java and Python~\cite{husain2019codesearchnet} and on the Wikipedia dataset~\cite{tensorflowWikipediaDataset}. All models used an embedding size of 500 and were trained for 20 epochs.

\textit{Testbeds.} Our evaluation spans eight system testbeds, seven obtained from public sources, and one collected internally by an intern. \cisco is a proprietary Cisco Systems dataset, not a public benchmark. \cisco was obtained through a confidential research collaboration, and its ground-truth trace links were created by actual Cisco engineers linking real pull requests to the production source files they modified. Because the underlying pull requests and source files are confidential, we report only aggregated, derived measurements, available in our online appendix~\cite{RepoTraceXplainer24}. \libest~\cite{libest} is an open-source project maintained also by Cisco, contains traceability links between requirements and test cases written in C (req2tc). \itrust, \etour, and \smos are widely adopted CoEST datasets that target use-case-to-source-code traceability (uc2src). The Event-Based Traceability corpus, \ebt, links English-language requirements to Java implementations, while \albergate concerns traceability among Java classes in a hotel management system. Finally, \dronology captures requirements-to-source-code links for an Unmanned Aerial System~\cite{dronology}. A summary of all datasets, including artifact counts and the number of traceability links, is provided in \tabref{tab:datasets}.

%%% TABLE TESTBEDS
%\input{tables/systems_dataset}
\begin{table}[t]
\centering
\caption{System Testbeds used in the Empirical Evaluation.}
\label{tab:datasets}
\vspace{-0.5em}
\scalebox{0.85}{
\setlength{\tabcolsep}{7pt}
\renewcommand{\arraystretch}{0.95}
\footnotesize
\begin{tabular}{llllrrr}
\toprule
 & & & & \multicolumn{3}{c}{\textbf{Datapoints}} \\
\cmidrule(lr){5-7}
\textbf{ID} & \textbf{System} & \textbf{Lang.} & \textbf{Link} & \textbf{All} & \textbf{Links} & \textbf{Non-L.} \\
\midrule
\stlibest{}     & \libest{}~\cite{libest}        & EN & req2tc  & 1{,}092  & 352     & 740      \\
\stcisco{}      & \cisco{}                       & EN & pr2src  & 21{,}312 & 547     & 20{,}765 \\
\stalbergate{}  & \albergate{}~\cite{1041053}    & EN & req2src & 935      & 53      & 882      \\
\stebt{}        & \ebt{}~\cite{1232285}          & EN & req2src & 2{,}050  & 98      & 1{,}952  \\
\stetour{}      & \etour{}~\cite{CoESTDatasets}  & IT & uc2src  & 6{,}728  & 308     & 6{,}420  \\
\stitrust{}     & \itrust{}~\cite{coestWebsite}  & EN & uc2src  & 47{,}815 & 277     & 47{,}538 \\
\stsmos{}       & \smos{}~\cite{6080780}         & IT & uc2src  & 6{,}700  & 1{,}044 & 5{,}656  \\
\stdronology{}  & \dronology{}~\cite{dronology}  & EN & req2src & 10{,}672 & 393     & 10{,}279 \\
\bottomrule
\end{tabular}
}
\vspace{-2em}
\end{table}

%%% TABLE TESTBEDS

\textit{Experimental Setup.} \tabref{tab:performance} summarizes our experiments in three dimensions: the evaluation system testbed, the model pretraining parameters, and the IR link type. Each configuration is evaluated using both vectorization techniques \wv and \dv. Our baseline experiment, \exbase, uses a model pretrained in Java and Python sources with conventional preprocessing to recover requirement links. The remaining experiments each introduce a single perturbation and measure its effect on the entropy level. Specifically, \exfirst and \exsecond vary the preprocessing tokenizer, using bpe8k and bpe32k, respectively. \exthird retains conventional preprocessing but replaces the pretraining corpus with Wikipedia. Finally, \exfourth varies the evaluation testbeds while keeping the same model as \exbase.

\textit{Metrics.} To address the formulated research questions, we leveraged the previously described datasets, which cover traceability corpora drawn from both industry and academic research, and introduced unsupervised machine learning models to predict traceability links. Each corpus contains multiple types of software artifacts as sources and targets; including requirements, test cases, and source code, together with their ground-truth links. The various information-theoretic metrics (\ie informativeness) were computed using DIT~\cite{dit}, a widely used Python library for discrete information theory. Our approach uses the information content of the source artifacts (\eg requirements, pull requests, use cases) to match them against target artifacts (\ie code). The following sub-sections describe the specific metrics used to answer each research question.
%%% TABLE Experiments
%\input{tables/models}

%\input{tables/tab_industry_experiments.tex} Removed and merged with the AUC performance table
%%% TABLE Experiments

\subsection{\ref{rq:effectivess}: Neural Unsupervised Effectiveness} 
To evaluate the effectiveness of neural unsupervised techniques for traceability, we computed the \textit{area under the curve (AUC)} from precision-recall curves and the \textit{receiver operating characteristic curve (ROC)} using the \textit{scikit-learn} API~\cite{sklearn_api}, similar to previous work that evaluated automated traceability techniques~\cite{moran_improving_2020,Dit2013b,Dasgupta2013,Guo2013FoundationsTraceability}.

\subsection{\ref{rq:semantic}: Semantic Traceability Imbalance} To answer~\ref{rq:semantic}, we estimated the following semantic metrics, each computed under both neural vectorization techniques (\ie \wv and \dv): Euclidean distance (\euc), Soft Cosine Similarity (\scm), Cosine Distance (\cosine), and Word Mover's Distance (\wmd). For \cosine and \wmd, we additionally computed their normalized inverses to obtain the corresponding \cosine and \wmd similarities. All distances and similarities were calculated over source-target artifact pairs in each testbed, where each pair is represented in the vector space produced by the unsupervised encoder. By segregating these values according to their ground-truth labels (\ie, actual links vs.\ non-links), we can reveal imbalances in the semantic space, as illustrated in \figref{fig:tracexplainer}.

\subsection{\ref{rq:exploratory}: Exploratory Information Analysis}  

To compute the set of information measures introduced in \approach, we conducted an \textit{Exploratory Data Analysis (EDA)}. Our \textit{EDA} consisted of an exhaustive statistical search for patterns and descriptive statistics over the reported information-theoretic measures (\ie informativeness), enabling us to interpret how well an unsupervised technique performs at traceability recovery. The goal is to use information measures to both describe and interpret the effectiveness of unsupervised traceability techniques. To this end, we proposed two complementary analyses.

\textbf{$AN_1$: \textit{Manifold of Information Measures.}} This analysis aims to characterize the probability distribution of each entropy and similarity metric. We hold prior assumptions about the expected shape of these distributions; for instance, similarity distributions should be bimodal, reflecting the two underlying populations of links and non-links. When the observed distribution deviates from these assumptions, we can use the discrepancy to assess the quality of the technique.

\textbf{$AN_2$: \textit{Manifold of Information Measures by Ground Truth.}} This analysis partitions each entropy and similarity metric according to the ground-truth labels. Segregating the data in this way allows us to interpret the prediction quality of the similarity metrics. It also allows us to describe how effectively the ground-truth signal was leveraged for each testbed, since \approach lets us monitor the information transmitted between source and target artifacts.

\subsection{\ref{rq:correlation}: Correlation Analyses} 
We correlate the semantic distance and similarity metrics with the information-theoretic measures using the \textit{Pearson coefficient}, and visualize the resulting relationships in a scatter matrix. The goal of this analysis is to surface strongly correlated pairs of unsupervised distances and information measures, which in turn help explain the rationale behind the traceability outputs in terms of the information-transmission setup proposed in \approach.

\section{Results \& Discussion}\label{sec:results}

%%TABLES
\begin{table}[t]
\centering
\caption{Experimental Setup and AUC Performance}
\label{tab:performance}
\vspace{-0.5em}
\scalebox{0.76}{
\setlength{\tabcolsep}{5pt}
\begin{tabular}{clccccccccc}
\toprule
\textbf{Exp.} & \textbf{Prep.} & \multicolumn{1}{c}{\textbf{System}} & \multicolumn{2}{c}{\textbf{W2V (WMD)}} & \multicolumn{2}{c}{\textbf{W2V (SCM)}} & \multicolumn{2}{c}{\textbf{D2V (COS)}} & \multicolumn{2}{c}{\textbf{D2V (EUC)}} \\
\textit{\textbf{ID}} & & \multicolumn{1}{c}{\textit{\textbf{Testbed}}} & \textit{\textbf{AUC}} & \textit{\textbf{ROC}} & \textit{\textbf{AUC}} & \textit{\textbf{ROC}} & \textit{\textbf{AUC}} & \textit{\textbf{ROC}} & \textit{\textbf{AUC}} & \textit{\textbf{ROC}} \\ \midrule
\multirow{3}{*}{\exbase} & \multirow{3}{*}{CONV} & \stlibest & 0.42 & 0.6 & 0.38 & 0.57 & 0.33 & 0.51 & 0.28 & 0.42 \\
 & & \stcisco& 0.04 & 0.62 & 0.04 & 0.62 & 0.05 & 0.64 & 0.02 & 0.4 \\
 & & \stdronology & 0.13 & 0.7& 0.17 & 0.75 & 0.09 & 0.65 & 0.04 & 0.48 \\ \hline
\multirow{3}{*}{\exfirst} & \multirow{3}{*}{BPE8} & \stlibest & 0.35 & 0.54 & 0.34 & 0.53 & 0.35 & 0.53 & 0.37 & 0.56 \\
 & & \stcisco& 0.03 & 0.54 & 0.03 & 0.5 & 0.03 & 0.51 & 0.02 & 0.49 \\
 & & \stdronology & 0.13 & 0.72 & 0.14 & 0.72 & 0.09 & 0.65 & 0.03 & 0.44 \\ \hline
\multirow{3}{*}{\exsecond} & \multirow{3}{*}{BPE32} & \stlibest & 0.34 & 0.53 & 0.3 & 0.49 & 0.31 & 0.49 & 0.32 & 0.5 \\
 & & \stcisco& 0.03 & 0.53 & 0.03 & 0.51 & 0.02 & 0.5 & 0.03 & 0.5 \\
 & & \stdronology & 0.13 & 0.71 & 0.13 & 0.72 & 0.09 & 0.68 & 0.03 & 0.44 \\ \hline
\multirow{3}{*}{\exthird} & \multirow{3}{*}{CONV} & \stlibest & 0.59 & 0.41 & 0.56 & 0.38 & 0.4 & 0.59 & 0.31 & 0.46 \\
 & & \stcisco& 0.05 & 0.64 & 0.05 & 0.66 & 0.04 & 0.61 & 0.02 & 0.4 \\
 & & \stdronology & 0.12 & 0.67 & 0.17 & 0.75 & 0.1 & 0.7 & 0.03 & 0.43 \\ \hline
\multirow{5}{*}{\exfourth} & \multirow{5}{*}{CONV} & \stalbergate & 0.08 & 0.58 & 0.15 & 0.77 & 0.09 & 0.53 & 0.05 & 0.49 \\
 & & \stebt & 0.2 & 0.81 & 0.14 & 0.79 & 0.18 & 0.73 & 0.04 & 0.44 \\
 & & \stetour & 0.23 & 0.75 & 0.24 & 0.74 & 0.15 & 0.72 & 0.04 & 0.46 \\
 & & \stitrust & 0.06 & 0.76 & 0.06 & 0.79 & 0.03 & 0.73 & 0 & 0.24 \\
 & & \stsmos & 0.2 & 0.57 & 0.21 & 0.6 & 0.17 & 0.54 & 0.12 & 0.41 \\ \bottomrule
\end{tabular}
}
\begin{tablenotes}
%\small
\item[*]{ \scriptsize\textit{AUC} is a precision-recall area under the curve. CONV: conventional tokenizer; BPE8/BPE32: SentencePiece BPE with 8k/32k vocabulary; \exbase - \exsecond, \exfourth are pretrained on Java \& Python;  \exthird is pretrained on Wikipedia + Java \& Python.}

\end{tablenotes}
\vspace{0.5em}
\end{table}

Our empirical evaluation comprises four experiments covering four data-science tasks: (i) describing distance and information measures, (ii) predicting trace links, (iii) estimating correlations between semantic distances and information measures, and (iv) characterizing information transmission. Each task is evaluated in three configuration axes: preprocessing strategy (conventional or BPE), embedding type (\wv or \dv) and pretraining dataset (CodeSearchNet or Wikipedia).

\subsection{\ref{rq:effectivess}: Traceability Effectiveness Results}

\begin{figure}[h]
   % Row 2
  \vspace{0.2cm}
  \begin{subfigure}{0.23\textwidth}
        \centering
        \includegraphics[width=\linewidth]{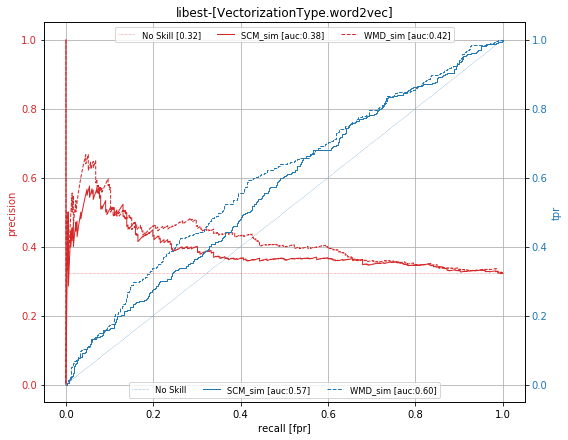}
        \caption{\tiny\libest \wv recall and precision}
        \label{fig:sub11}
    \end{subfigure}
    %\hfill
    \begin{subfigure}{0.23\textwidth}
        \centering
        \includegraphics[width=\linewidth]{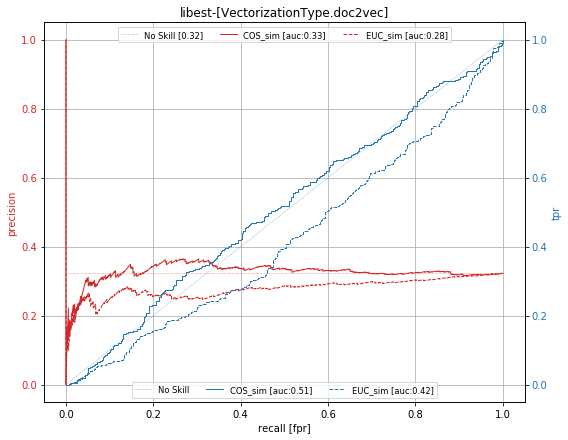}
        \caption{\tiny\libest \dv recall and precision}
        \label{fig:sub12}
    \end{subfigure}
    \vspace{0.2cm}
    \begin{subfigure}{0.23\textwidth}
        \centering
        \includegraphics[width=\linewidth]{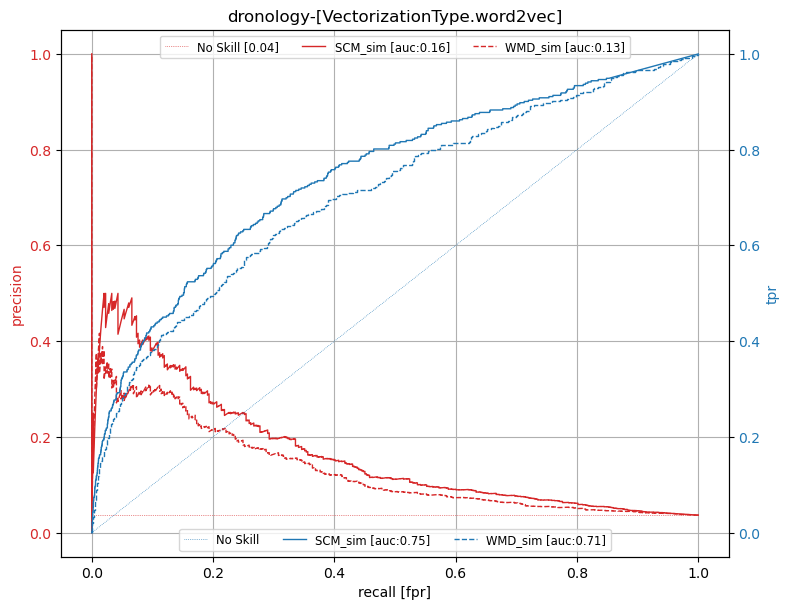}
        \caption{\tiny\dronology \wv recall and precision}
        \label{fig:sub23}
    \end{subfigure}
    %\hfill
    \begin{subfigure}{0.23\textwidth}
        \centering
        \includegraphics[width=\linewidth]{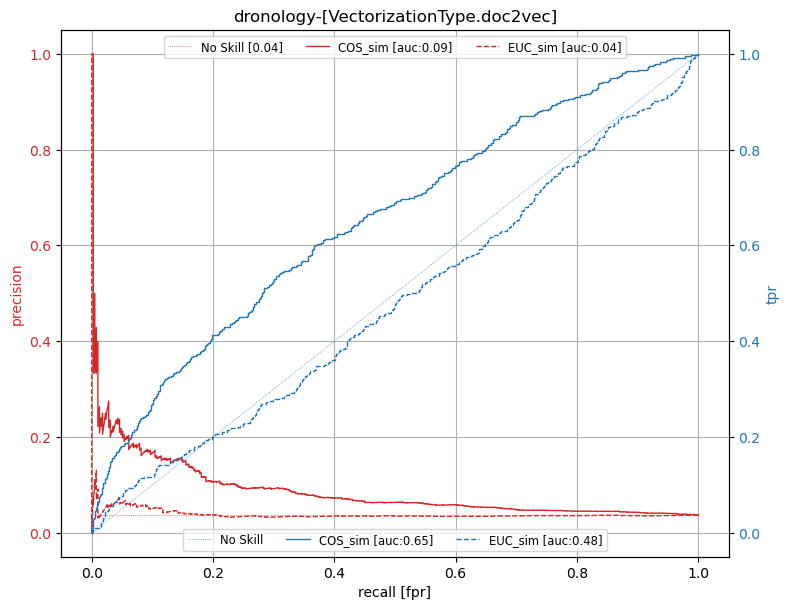}
        \caption{\tiny\dronology \dv recall and precison}
        \label{fig:sub24}
    \end{subfigure}
     \caption{Precision and Recall for \wv and \dv using \libest and \dronology}
    \label{fig:grid1}
\end{figure}

We evaluate traceability via link-recovery accuracy using precision, precision–recall AUC (PR-AUC), and ROC-AUC, where higher values indicate better performance. Table~\tabref{tab:performance} summarizes results for the unsupervised techniques (\ie \wv and \dv) across all experiments. None of the configurations achieves strong classification performance: PR-AUC remains below 0.6, and although ROC-AUC reaches up to 0.76, this overstates performance under our data conditions. Because true links constitute a small fraction of candidate pairs in every testbed (\eg 277 out of 47,815 for \itrust), ROC-AUC is inflated relative to PR-AUC. In contrast, PR-AUC is more sensitive to the rare positive class and therefore provides a more reliable assessment in this setting. Accordingly, we interpret the consistently low PR-AUC as the primary indicator of classifier effectiveness. Across all experiments, \wv representations (both WMD and SCM) consistently outperform \dv, most notably on \libest in \exthird, where PR-AUC reaches $0.56–-0.59$ and ROC-AUC peaks near 0.7, while \dv remains lower under the same conditions.

%We evaluate traceability through link-recovery accuracy, reported as precision, recall AUC, and ROC, with higher values indicating better performance. \tabref{tab:performance} reports both unsupervised techniques (\ie \wv and \dv) across all experiments. \revision{None of the configurations yields a strong classifier: AUC stays below $0.6$, and although ROC reaches up to $0.76$, this is not indicative of strong performance under our data conditions. Because ground-truth links are a small minority of all candidate pairs in every testbed (\eg $277$ of $47{,}815$ for \itrust), ROC-AUC is inflated relative to precision-recall AUC, precision-recall AUC is far more sensitive to the (rare) positive class and remains the more informative metric here. We therefore read the low precision-recall AUC, not the higher ROC, as the more reliable indicator of classifier strength.} \wv representations (both WMD and SCM) consistently outperform \dv, most notably on \libest in \exthird, where AUC reaches $0.56$--$0.59$ and ROC peaks near $0.7$, while \dv remains lower under the same setting.

%%%%% TABLES
\begin{table}[]
\centering

\caption{Self information, Loss, Mutual Information and Minimum Shared Information}
\label{tab:information}
%\vspace{-0.5em}
\scalebox{0.55}{
\setlength{\tabcolsep}{3.4pt} 
\begin{tabular}{cllrrrrlcccccccccc}
\toprule
\textbf{Exp.} & \multicolumn{1}{c}{\textbf{Sys.}} &  & \multicolumn{2}{c}{\textbf{Word2vec}} & \multicolumn{2}{c}{\textbf{Doc2Vec}} & \multicolumn{1}{c}{\textbf{}} & \textbf{SI S.} & \textbf{SI T.} &  & \textbf{CI Noise} &  & \textbf{CI Loss} &  & \textbf{MI} & \textbf{MSI} & \textbf{MSI} \\
\textit{\textbf{ID}} & \multicolumn{1}{c}{\textit{\textbf{ID}}} &  & \multicolumn{1}{l}{\textbf{SCM}} & \multicolumn{1}{l}{\textbf{WMD}} & \multicolumn{1}{l}{\textbf{COS}} & \multicolumn{1}{l}{\textbf{EUC}} &  & \textit{\textbf{H(X)}} & \textit{\textbf{H(Y)}} & \multirow{-2}{*}{\textbf{D1}} & \textit{\textbf{H(Y|X)}} & \multirow{-2}{*}{\textbf{D2}} & \textit{\textbf{H(X|Y)}} & \multirow{-2}{*}{\textbf{D3}} & \textit{\textbf{I(X:Y)}} & \textit{\textbf{Si[std]}} & \textit{\textbf{Sx[std]}} \\ \cline{1-2} \cline{4-7} \cline{9-18} 
\rule{0pt}{3ex}    
 & \stlibest &  & 0.28 & 0.49 & 0.17 & 0.01 &  & 5.54 & 7.77 & 2.23 & 0.18 & 7.59 & 2.4 & 3.14 & 5.37 & 3.82[0.66] & 1.37[0.03] \\
& \stcisco &  & \cellcolor[HTML]{FFCCC9}0.1 & 0.45 & 0.09 & 0.02 &  & 3.42 & 5.91 & \cellcolor[HTML]{DAE8FC}2.49 & 0.21 & 5.7 & 2.7 & \cellcolor[HTML]{DAE8FC}0.72 & 3.21 & 1.45[1.14] & 0.87[0.54] \\
\multirow{-3}{*}{\exbase} & \stdronology &  & 0.16 & 0.46 & 0.19 & 0.02 &  & 4.37 & 5.86 & 1.49 & 0.28 & 5.58 & 1.77 & 2.6 & 4.09 & 2.32[0.77] & 1.25[0.12] \\
\cline{1-2} \cline{4-7} \cline{9-18} 
\rule{0pt}{3ex}
 & \stlibest &  & 0.42 & 0.51 & 0 & \cellcolor[HTML]{FFCCC9}0.98 &  & 6.58 & 7.33 & 0.75 & 0.12 & 7.21 & 0.86 & 5.72 & 6.46 & \cellcolor[HTML]{DAE8FC}5.77[0.6] & 1.42[0.02] \\
 & \stcisco &  & 0.28 & 0.47 & 0 &\cellcolor[HTML]{FFCCC9} 0.98 &  & 4.68 & 6.6 & 1.92 & 0.17 & 6.43 & 2.08 & 2.6 & 4.52 & 3.17[1.44] & 1.24[0.35] \\
 \multirow{-3}{*}{\exfirst} & \stdronology &  & 0.09 & 0.46 & 0.2 & 0.02 &  & 4.12 & 5.67 & 1.55 & 0.28 & 5.39 & 1.83 & 2.29 & 3.84 & 1.98[0.69] & 1.20[0.13] \\
\cline{1-2} \cline{4-7} \cline{9-18} 
\rule{0pt}{3ex}
 & \stlibest &  & 0.37 & 0.49 & 0 &\cellcolor[HTML]{FFCCC9} 0.98 &  & 6.54 & 7.47 & 0.93 & 0.14 & 7.33 & 1.07 & 5.47 & 6.4 & 5.42[0.64] & 1.42[0.01] \\
& \stcisco &  & 0.18 & 0.46 & 0 & \cellcolor[HTML]{FFCCC9}0.98 &  & 4.42 & 6.56 & 2.14 & 0.26 & 6.3 & 2.4 & 2.02 & 4.16 & 2.63[1.41] & 1.15[0.41] \\
\multirow{-3}{*}{\exsecond} & \stdronology &  &  \cellcolor[HTML]{FFCCC9}0.07 & 0.44 & 0.2 & 0.02 &  & 3.48 & 5.23 & 1.75 & 0.28 & 4.95 & 2.03 & 1.45 & 3.2 & 1.43[0.53] & 1.08[0.13] \\
\cline{1-2} \cline{4-7} \cline{9-18} 
\rule{0pt}{3ex}

 & \stlibest &  & 0.23 & 0.49 & 0.26 & 0.01 &  & 5.54 & 7.77 & 2.23 & 0.18 & 7.59 & 2.4 & 3.14 & 5.37 & 2.82[0.66] & 1.37[0.03] \\
 & \stcisco &  & 0.08 & 0.45 & 0.17 & 0.02 &  & 3.42 & 5.91 & \cellcolor[HTML]{DAE8FC}2.49 & 0.21 & 5.7 & 2.7 & \cellcolor[HTML]{DAE8FC}0.72 & 3.21 & 1.45[1.14] & 0.87[0.54] \\
 \multirow{-3}{*}{\exthird} & \stdronology & &  \cellcolor[HTML]{FFCCC9} 0.07 & 0.44 & 0.2 & 0.02 & & 3.48 & 5.23 & 1.75 & 0.28 & 4.95 & 2.03 & 1.45 & 3.2 & 1.43[0.53] & 1.08[0.13] \\
\cline{1-2} \cline{4-7} \cline{9-18} 
\rule{0pt}{3ex}
 & \stalbergate &  & \cellcolor[HTML]{FFCCC9} 0.07 & 0.45 & 0.22 & 0.02 &  & 6.62 & 6.18 & 0.44 & 0.91 & 5.27 & 0.47 & 6.15 & 5.7 & 2.81[0.96] & 1.27[0.19] \\
 & \stebt &  & \cellcolor[HTML]{FFCCC9}0.1 & 0.45 & \cellcolor[HTML]{FFCCC9}0.1 & 0.03 &  & 2.97 & 4.73 & 1.76 & 0.25 & 4.48 & 2.01 & \cellcolor[HTML]{DAE8FC}0.96 & 2.72 & \cellcolor[HTML]{FFCCC9}0.61[0.72] & 0.5[0.54] \\
 & \stetour &  & 0.16 & 0.46 & 0.07 & 0.02 &  & 5.23 & 5.77 & 0.54 & 0.53 & 5.24 & 1.07 & 4.16 & 4.7 & 2.1[1.13] & 1.02[0.34] \\
 & \stitrust &  & 0.14 & 0.46 & 0.09 & 0.02 &  & 3.93 & 5.56 & 1.63 & 0.33 & 5.23 & 1.96 & 1.97 & 3.6 & 1.33[1.02] & 0.85[0.52] \\
\multirow{-5}{*}{\exfourth} & \stsmos &  & 0.05 & 0.45 & 0.13 & 0.02 &  & 5.09 & 5.7 & 0.61 & 0.55 & 5.15 & 1.16 & 3.93 & 4.54 & 1.44[0.71] & 1.02[0.34] \\
\bottomrule
\end{tabular}
}

{\scriptsize\textit{D1: $H(X)-H(Y)$, D2: $H(Y)-H(Y|X)$, D3: $H(X)-H(X|Y)$, S.: Source, T.: Target. Cells in red indicate the lowest values on each metric, while cells in blue are the highests}
}
\vspace{0.5em}
\end{table}
%%%%%

Performance varies notably between systems and experiments. \libest is almost always the top performer, followed by \dronology, while \cisco is consistently the weakest in \exbase--\exthird. In \exfourth, the smaller systems \ebt and \smos outperform \albergate and \itrust. As \tabref{tab:performance} shows, the data lack the patterns needed for reliable binary classification (\ie, link vs.\ non-link), and \dv in particular fails to retrieve links; confirming its ineffectiveness for this task.

\figref{fig:sub11} and \figref{fig:sub12} show precision and recall for \scm and \wmd on \libest under \wv and \dv: precision falls to $3.8$ (\scm{}) and $0.42$ (\wmd) for \wv, and \dv (\figref{fig:sub12}) performs poorly overall. By contrast, \figref{fig:sub23} and \figref{fig:sub24} show that on \dronology under \wv, both \wmd and \scm reach higher ROC ($0.71$, $0.75$) and precision–recall ($0.13$, $0.16$) areas, indicating better discriminative performance across the full operating range.

%\textbf{Summary.} The effectiveness observed on the AUC precision-recall for all experiments exhibits low performance when recovering the links at both vectorization types \dv and \wv with a slight improvement for \wv. 

\vspace{-0.5em}
\begin{boxK}
\vspace{-0.5em}
    \ref{rq:effectivess} Across all experiments, the AUC precision-recall results show consistently low link-recovery performance for both \dv and \wv, with \wv only marginally ahead.
\vspace{-1.5em}
\end{boxK}
\vspace{-0.2em}

\subsection{\ref{rq:semantic} Semantic Traceability Imbalance Results}

\tabref{tab:information} summarizes the observed results for soft-cosine similarity (\scm), Word Mover's Distance (\wmd), cosine distance (\cosine), and Euclidean distance (\euc). Low \scm values; for example, $0.1$ on both \cisco and \ebt, indicate weak similarity between source and target artifacts. The highest \wmd value, $0.51$ on \libest in \exfirst, suggests that \wmd is capturing meaningful token overlap between sources and targets. \tabref{tab:by_links} further breaks down these semantic metrics by links and non-links. Notably, our models retain low \scm values even for ground-truth links, signaling unrelated information between source and target; for instance, \smos reports an \scm of $0.06$ for confirmed links, indicating that the neural vectorization technique fails to recover semantic relatedness. Similarly, the $0.51$ \wmd for \libest in \exfirst reflects an insufficient number of related tokens between source and target. \dv exhibits comparable behavior, with Euclidean distances of $0.98$ on \libest and \cisco in \exfirst and \exsecond (see \tabref{tab:information}). In contrast, \dronology yields much lower Euclidean distances ($0.02$) in these experiments, consistent with its values in \exbase and \exthird. This gap likely reflects the larger and more diverse \dronology corpus, which preserves meaningful subword overlap under BPE preprocessing and thus avoids the near-maximal separation seen in the smaller datasets.

%%Introduce image here
%\textbf{Summary.} The ground truth for traceability suffers from extremely imbalanced link classes (\ie actual links and non-links). In addition, cosine (\cosine) distance and Soft-Cosine similarity (\scm) behave better under AUC analysis, indicating the unsupervised models identify links with a minimum effectiveness of 0.19 and 0.12, respectively, for \cisco testbed. .

\vspace{-0.5em}
\begin{boxK}
\vspace{-0.5em}
    \ref{rq:semantic} The traceability ground truth is heavily imbalanced between links and non-links. \cosine and \scm fare best under AUC, with minimum effectiveness of $0.19$ and $0.12$ on \cisco.
\vspace{-1.5em}
\end{boxK}
\vspace{-0.2em}

\subsection{\ref{rq:exploratory} Exploratory Information Theory Results}

%%%%% TABLES

\begin{table*}[]
\centering

\caption{Self information, Loss, Mutual Information and Minimum Shared Information by Links}
\label{tab:by_links}
%\vspace{-0.5em}
\scalebox{0.8}{

\setlength{\tabcolsep}{3pt} 
\begin{tabular}{cllllllllllllllllllllllllll}
\toprule
\multicolumn{1}{l}{} &  &  & \multicolumn{2}{c}{\textbf{Similarity}} &  & \multicolumn{6}{c}{\textbf{Distance}} &  & \multicolumn{14}{c}{\textbf{Information}} \\
\textbf{Exp.} & \multicolumn{1}{c}{\textbf{Sys.}} &  & \multicolumn{2}{c}{\textbf{SCM}} &  & \multicolumn{2}{l}{\textbf{WMD}} & \multicolumn{2}{c}{\textbf{COS}} & \multicolumn{2}{c}{\textbf{EUC}} &  & \multicolumn{2}{c}{\textbf{H(X)}} & \multicolumn{2}{c}{\textbf{H(Y)}} & \multicolumn{2}{c}{\textbf{H(Y|X)}} & \multicolumn{2}{c}{\textbf{H(X|Y)}} & \multicolumn{2}{c}{\textbf{I(X:Y)}} & \multicolumn{2}{c}{\textbf{Si(X:Y)}} & \multicolumn{2}{c}{\textbf{Sx(X:Y)}} \\
\textit{\textbf{ID}} & \multicolumn{1}{c}{\textit{\textbf{ID}}} &  & \multicolumn{1}{c}{\textit{\textbf{Link}}} & \multicolumn{1}{c}{\textit{\textbf{NoL}}} &  & \multicolumn{1}{c}{\textit{\textbf{Link}}} & \multicolumn{1}{c}{\textit{\textbf{NoL}}} & \multicolumn{1}{c}{\textit{\textbf{Link}}} & \multicolumn{1}{c}{\textit{\textbf{NoL}}} & \multicolumn{1}{c}{\textit{\textbf{Link}}} & \multicolumn{1}{c}{\textit{\textbf{NoL}}} &  & \multicolumn{1}{c}{\textit{\textbf{Link}}} & \multicolumn{1}{c}{\textit{\textbf{NoL}}} & \multicolumn{1}{c}{\textit{\textbf{Link}}} & \multicolumn{1}{c}{\textit{\textbf{NoL}}} & \multicolumn{1}{c}{\textit{\textbf{Link}}} & \multicolumn{1}{c}{\textit{\textbf{NoL}}} & \multicolumn{1}{c}{\textit{\textbf{Link}}} & \multicolumn{1}{c}{\textit{\textbf{NoL}}} & \multicolumn{1}{c}{\textit{\textbf{Link}}} & \multicolumn{1}{c}{\textit{\textbf{NoL}}} & \multicolumn{1}{c}{\textit{\textbf{Link[std]}}} & \multicolumn{1}{c}{\textit{\textbf{NoL[std]}}} & \multicolumn{1}{c}{\textit{\textbf{Link[std]}}} & \multicolumn{1}{c}{\textit{\textbf{NoL[std]}}} \\ \cline{1-2} \cline{4-5} \cline{7-12} \cline{14-27} 
\rule{0pt}{3ex}   
 & \stlibest &  & 0.30 & 0.27 &  & 0.49 & 0.48 & 0.17 & 0.17 & 0.01 & 0.01 &  & 5.70 & 0.00 & 7.73 & 7.78 & 0.19 & 0.17 & 2.22 & 2.48 & 5.51 & 5.30 & 3.95[0.7] & 3.76[0.63] & 1.37[0.04] & 1.37[0.03] \\
& \stcisco &  & \cellcolor[HTML]{FFCCC9}0.14 & 0.10 &  & 0.46 & 0.45 & 0.13 & 0.09 & 0.02 & 0.02 &  & 3.80 & 3.41 & 6.23 & 5.90 & 0.20 & 0.21 & 2.63 & 2.71 & 3.60 & 3.20 & 1.98[1.11] & 1.44[1.14] & 10.7[0.43] & 0.86[0.54] \\
\multirow{-3}{*}{\exbase} & \stdronology & & 0.18 & 0.07 & & 0.47 & 0.45 & 0.26 & 0.17 & 0.06 & 0.06 &  & 3.53 & 3.47 & 5.4 & 5.23 & 0.21 & 0.28 & 2.08 & 2.03 & 3.32 & 3.19 & 1.77[0.61] & 1.41[0.52] & 1.16[0.13] & 1.08[0.13] \\
\cline{1-2} \cline{4-5} \cline{7-12} \cline{14-27} 
\rule{0pt}{3ex}   
 & \stlibest &  & 0.43 & 0.42 &  & \cellcolor[HTML]{FFCCC9}0.51 & 0.51 & 0.00 & 0.00 & 0.98 & 0.98 &  & 6.80 & 6.48 & 7.31 & 7.33 & 0.13 & 0.11 & 0.64 & 0.97 & 6.67 & 6.37 & 5.94[0.53] & 6.68[0.062] & 1.42[0.01] & 1.42[0.02] \\
& \stcisco &  & 0.28 & 0.28 &  & 0.47 & 0.47 & 0.00 & 0.00 & 0.98 & 0.98 &  & 4.94 & 4.68 & 6.72 & 6.59 & 0.16 & 0.17 & 1.93 & 2.08 & 4.79 & 4.51 & 3.59[1.26] & 3.16[1.44] & 1.31[0.26] & 1.24[0.35] \\
\multirow{-3}{*}{\exfirst} & \stdronology & & 0.26 & 0.15 & & 0.47 & 0.46 & 0.23 & 0.19 & 0.02 & 0.02 & & 4.50 & 4.36 & 6 & 5.85 & 0.22 & 0.28 & 1.73 & 1.76 & 4.28 & 4.08 & 2.86[0.71] & 2.29[0.76] & 1.31[0.08] & 1.24[0.12] \\
\cline{1-2} \cline{4-5} \cline{7-12} \cline{14-27} 
\rule{0pt}{3ex}   
 & \stlibest &  & 0.37 & 0.37 &  & 0.49 & 0.49 & 0.00 & 0.00 & 0.98 & 0.98 &  & 6.71 & 6.46 & 7.44 & 7.49 & 0.15 & 0.13 & 0.88 & 1.16 & 6.56 & 6.33 & 5.58[0.57] & 5.35[0.65] & 1.42[0.01] & 1.41[0.01] \\
 & \stcisco &  & 0.18 & 0.18 &  & 0.46 & 0.46 & 0.00 & 0.00 & 0.98 & 0.98 &  & 4.66 & 4.41 & 6.71 & 6.56 & 0.26 & 0.26 & 2.31 & 2.41 & 4.40 & 4.15 & 3.04[1.26] & 2.62[1.41] & 1.25[0.31] & 1.15[0.41] \\
 \multirow{-2}{*}{\exsecond} & \stdronology & & 0.21 & 0.12 & & 0.47 & 0.46 & 0.25 & 0.20 & 0.02 & 0.02 & & 4.25 & 4.11 & 5.81 & 5.66 & 0.23 & 0.28 & 1.79 & 1.83 & 4.02 & 3.83 & 2.46[0.72] & 1.95[0.68] & 1.27[0.1] & 1.19[0.13] \\
\cline{1-2} \cline{4-5} \cline{7-12} \cline{14-27} 
\rule{0pt}{3ex}   
 & \stlibest &  & 0.25 & 0.23 &  & 0.49 & 0.48 & \cellcolor[HTML]{DAE8FC}0.28 & \cellcolor[HTML]{DAE8FC}0.26 & 0.01 & \cellcolor[HTML]{FFCCC9}0.01 &  & 5.70 & 5.47 & 7.73 & 7.78 & 0.19 & 0.17 & 2.22 & 2.48 & 5.51 & 5.30 & 3.95[0.7] & 3.76[0.63] & 1.37[0.04] & 1.37[0.03] \\
& \stcisco &  & \cellcolor[HTML]{FFCCC9}0.12 & 0.08 &  & 0.46 & 0.45 & 0.19 & 0.16 & 0.01 & 0.02 &  & 3.80 & 3.41 & 6.23 & 5.90 & 0.20 & 0.21 & 2.63 & 2.71 & 3.60 & 3.20 & 1.98[1.11] & 1.44[1.14] & 1.07[0.43] & 0.86[0.54] \\
\multirow{-2}{*}{\exthird} & \stdronology & & 0.18 & 0.07 & & 0.46 & 0.44 & 0.26 & 0.20 & 0.02 & 0.02 & & 3.53 & 3.48 & 5.4 & 5.23 & 0.21 & 0.28 & 2.08 & 2.03 & 3.32 & 3.19 & 1.77[0.61] & 1.41[0.52] & 1.16[0.13] & 1.08[0.13] \\
\cline{1-2} \cline{4-5} \cline{7-12} \cline{14-27} 
\rule{0pt}{3ex}   
 & \stalbergate &  & \cellcolor[HTML]{FFCCC9}0.11 & 0.07 &  & 0.46 & 0.45 & 0.23 & 0.22 & 0.02 & 0.02 &  & 6.70 & 6.61 & 6.19 & 6.18 & 0.93 & 0.91 & 0.42 & 0.48 & 5.77 & 5.70 & 3.07[0.91] & 2.8[0.96] & 1.3[0.1] & 1.27[0.2] \\
 & \stebt &  & 0.20 & 0.09 &  & 0.47 & 0.45 & 0.16 & 0.10 & 0.03 & 0.03 &  & 2.94 & 2.97 & 5.11 & 4.71 & 0.15 & 0.26 & 2.32 & 2.00 & 2.78 & 2.71 & 1.2[0.74] & 0.56[0.7] & 0.9[0.46] & 0.46[0.53] \\
 & \stetour &  & 0.24 & 0.15 &  & 0.47 & 0.46 & 0.11 & 0.07 & 0.02 & 0.02 &  & 5.28 & 5.22 & 5.79 & 5.76 & 0.49 & 0.53 & 1.00 & 1.07 & 4.79 & 4.69 & 2.48[0.91] & 2.08[1.03] & 1.22[0.2] & 1.13[0.35] \\
 & \stitrust &  & 0.28 & 0.14 &  & 0.48 & 0.46 & 0.15 & 0.09 & 0.01 & 0.02 &  & 4.07 & 3.92 & 6.28 & 5.55 & 0.12 & 0.33 & 2.33 & 1.95 & 3.95 & 3.60 & 2.51[1.07] & 1.32[1.01] & 1.21[0.27] & 0.85[0.52] \\
\multirow{-5}{*}{\exfourth} & \stsmos &  & \cellcolor[HTML]{FFCCC9}0.06 & 0.05 &  & 0.45 & 0.45 & 0.13 & 0.13 & 0.02 & 0.02 &  & 5.08 & 5.09 & 5.82 & 5.67 & 0.45 & 0.57 & 1.19 & 1.15 & 4.63 & 4.52 & 1.62[0.66] & 1.41[0.71] & 1.1[0.2] & 1[0.36] \\
\bottomrule
\end{tabular}

}
{
\vspace{0.2em}
\\
\scriptsize\textit{Cells highlighted in red mark the lowest value observed for each metric, while cells highlighted in blue mark the highest.}
}
\vspace{-1
em}
\end{table*}
%%%%%

$AN_1$: \textit{Information measures.} \tabref{tab:information} depicts the self-information for source and target artifacts, the noise, loss, mutual information, and the minimum shared information. We observe the self-information $H(X)$ of the source artifacts (or issues) in on average $4.63[1.17]$\bits, while the self-information of the target artifacts (or source code) $H(Y)$ is on average $6.16[0.91]$\bits suggesting that the amount of information of $H(Y)$ is on average 1.53\bits larger than the amount of information in the set of $H(X)$. The last means that the source has less information than the target demonstrating information imbalance.

We observe the highest noise at the experiment \exfourth with \albergate with a value of 0.91\bits. In particular, when computing the difference between self-information $H(X)$ and $H(Y)$, we observe the highest unbalanced information for the \cisco system in the experiments \exbase and \exthird with a value of 2.49\bits. This result indicates non-related information between the source and target and confirms the semantic distance results \scm for the same testbeds and experiments.

The Mutual Information averages $4.33[1.14]$\bits indicating a positive transmission of information from the source to the target. The Minimum Share Information Entropy $Si$ is on average $2.43[1.39]$\bits, and the Minimum Shared Extropy $Sx$ is on average $1.12[0.24]$\bits for all experiments suggesting the source and target possess information in common. However, the target could not share information with the source as depicted at \exfourth with \ebt testbed where we observe an $Si$ of 0.61\bits and a loss of 2.01\bits. The loss difference of 0.96\bits and similarity of 0.1\bits demonstrates an information imbalance between the source and target. In contrast, the $Si$ value for \libest at the \exfirst reports the highest value with 5.77\bits. In contrast, the loss and noise differences are high with 5.72\bits and 7.21\bits respectively.

The loss and noise are Gaussian distributions with a median of 1.82\bits and 0.30\bits respectively. The loss is larger than the noise by a range of 1.52\bits on average for all the experiments. When the estimated median of the minimum shared entropy is 2.43\bits, we detected a high amount of information, suggesting related information but poorly commented source code. Furthermore, the noise is barely a bit unit, indicating that the code is not influenced by an external source of information.

The distance and similarity metrics reflect a non-standard behavior, for instance, when comparing the cosine (for \dv) and the \wmd (for \wv). The average value for conventional preprocessing (\ie \exbase and \exfourth) of the cosine is $0.14[0.05]$, while the value for the \wmd is $0.46[0.02]$. Both distance distributions are unimodal entailing the binary classification does not exist and both distance metrics do not overlap. 

%\textbf{Summary.} The maximum information transmission is around 4.52\bits from issues to source code for \cisco testbed at the \exfirst. We recommend that software developers implement inspection procedures to refactor documentation in both requirements and source code to enhance mutual information (and MSI). 

\vspace{-0.5em}
\begin{boxK}
%\vspace{-0.5em}
    \ref{rq:exploratory}-$AN_1$ On the \cisco testbed under \exfirst, information transmission from issues to code peaks at $\approx 4.52\,\bits$. To raise \textit{MI} and \textit{MSI}, we recommend develop teams adopt inspection procedures that refactor both requirement and code documentation.
%\vspace{-1.5em}
\end{boxK}
\vspace{-0.2em}

%If we observe the link ${PR-294 \to psb\_mapping.py}$, which corresponds to the minimum MI of 5.5\bits the 99\% quantile,   then we infer that the 4.4\bits of maximum transmission can be improved until reaching an average value of 5.5\bits.  

$AN_2$: \textit{Information measures by Ground Truth.} Unfortunately, the information measures are unaffected by the nature of the traceability relationship (see \tabref{tab:by_links}): $H(X)$ and $H(Y)$ are on average comparable for links and non-links, meaning that per-artifact information is independent of link existence. Although sequence-based artifacts share at least some information, this \textit{independence} should not be transferred to similarity metrics such as \scm, \euc, or \wmd. Consequently, skip-gram-based neural unsupervised techniques cannot reliably perform binary link classification; the data simply do not encode the necessary discriminative patterns. Closing this gap requires probabilistic models that intervene in the expected value of a link~\cite{moran_improving_2020} or systematic refactorings of the artifacts themselves.

%\revision{This null result admits at least three non-exclusive explanations that our design cannot disentangle: (1) \textit{model failure}, where the semantic-distance metrics (\scm, \euc, \cosine, \wmd) do not separate links from non-links; (2) \textit{artifact information gap}, where valid links share too little overlapping information for MI to detect (\eg poor described PRs, under-documented code); and (3) \textit{ground-truth noise}, where labels are incomplete or incorrect, pulling the MI of  ``confirmed'' links toward non-links (\eg $Case_3$, \secref{sec:cases}). Resolving these requires auditing disputed links with independent raters and reporting agreement which is beyond the scope of this work.}

%%PASTE HERE A TABLE

%\textbf{Summary.} Although the information amount in the source code is larger than in the set of issues; the MI, loss, and noise are indistinguishable from confirmed links to non-links. We expect low mutual information values and high loss and noise information amounts for non-related artifacts. \approach's information-theory measures indicate when and why unsupervised models fail, thereby making their effectiveness and limitations more transparent and data-driven.

\vspace{-0.5em}
\begin{boxK}
%\vspace{-0.5em}
    \ref{rq:exploratory}-$AN_2$ Although code carries more information than the corresponding issues, the MI, loss, and noise values are indistinguishable between confirmed links and non-links; even though unrelated artifacts should exhibit low MI together with high loss and noise. By exposing when and why unsupervised models fail, \approach's information-theoretic measures make their effectiveness and limitations more transparent and data-driven.
%\vspace{-1.5em}
\end{boxK}
\vspace{-0.2em}

\subsection{\ref{rq:correlation} Correlation Results}

\begin{figure}[ht]
    \centering
    % Row 1
    \begin{subfigure}{0.23\textwidth}
        \centering
        \includegraphics[width=\linewidth]{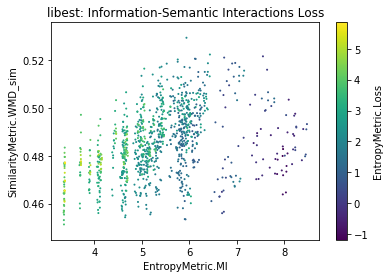}
        \caption{\tiny\libest \wmd similarity, MI and Loss }
        \label{fig:sub51}
    \end{subfigure}
    %\hfill
    \begin{subfigure}{0.23\textwidth}
        \centering
        \includegraphics[width=\linewidth]{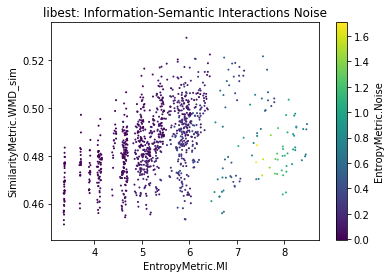}
        \caption{\tiny\libest \wmd similarity, MI and Noise}
        \label{fig:sub52}
    \end{subfigure}
    \vspace{0.2cm}
    \begin{subfigure}{0.23\textwidth}
        \centering
        \includegraphics[width=\linewidth]{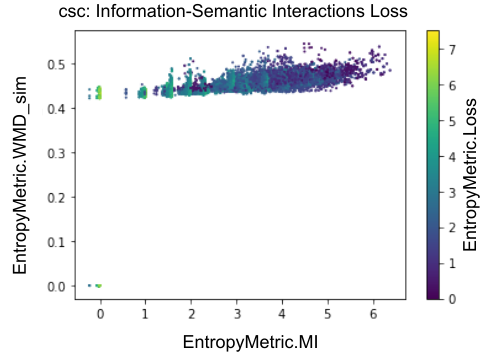}
        \caption{\tiny\cisco \wmd similarity, MI and Loss}
        \label{fig:sub53}
    \end{subfigure}
    %\hfill
    \begin{subfigure}{0.23\textwidth}
        \centering
        \includegraphics[width=\linewidth]{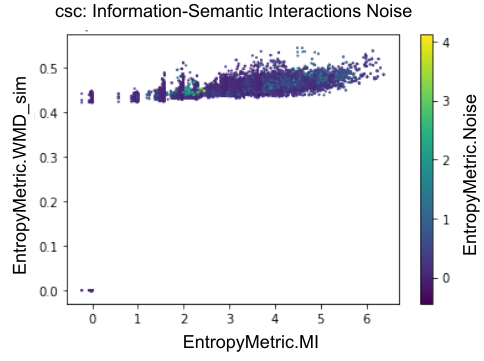}
        \caption{\tiny\cisco \wmd similarity, MI and Noise}
        \label{fig:sub54}
    \end{subfigure}\vspace{0.2cm}
    \begin{subfigure}{0.23\textwidth}
        \centering
        \includegraphics[width=\linewidth]{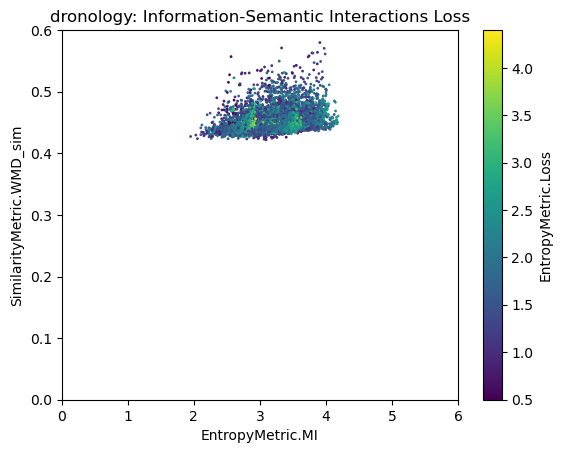}
        \caption{\tiny\dronology \wmd similarity, MI and Loss}
        \label{fig:sub55}
    \end{subfigure}
    %\hfill
    \begin{subfigure}{0.23\textwidth}
        \centering
        \includegraphics[width=\linewidth]{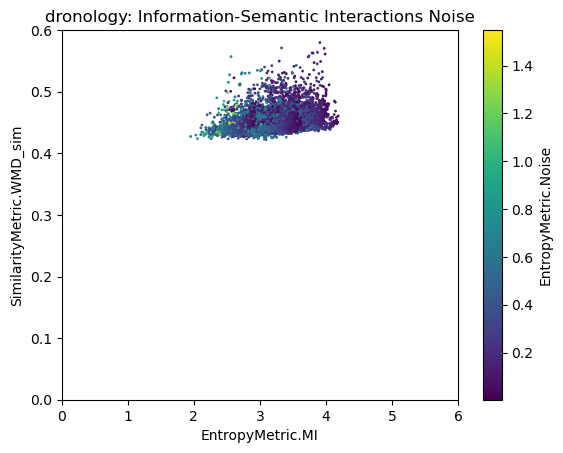}
        \caption{\tiny\cisco \dronology similarity, MI and Noise}
        \label{fig:sub56}
    \end{subfigure}
   
\vspace{-0.2cm}
    \caption{Similarity Vs Mutual Information Vs Noise and Loss for \libest, \cisco and \dronology.}
    \label{fig:grid2}
    \vspace{-0.2cm}
\end{figure}

\textit{Scatter Matrix for Information Measures.} 
Correlation analysis exposes structure that is not visible from raw values, helping us identify shared patterns or causal cues across metrics. We therefore correlate the similarity variables with one another and with the information measures (\eg MI, loss, noise, entropy). As shown in \tabref{tab:correlations}, \wmd similarity is predominantly positively correlated ($\approx 0.74$, \figref{fig:sub53}) with the information metrics, whereas \cosine similarity displays the opposite behavior.

\begin{table}[]
\centering
\caption{Correlation between the distance and entropy}
\label{tab:correlations}
\vspace{-0.5em}
\scalebox{0.72}{

\setlength{\tabcolsep}{3pt} 
\begin{tabular}{cccccccccccc}
\textbf{Exp.} & \textbf{Sys.} & \multicolumn{2}{c}{\textbf{Mutual Information}} & \multicolumn{2}{c}{\textbf{Loss}} & \multicolumn{2}{c}{\textbf{Noise}} & \multicolumn{4}{c}{\textbf{Minimum Shared Entropy}} \\
\textit{\textbf{ID}} & \textit{\textbf{ID}} & \textit{\textbf{WMD}} & \textit{\textbf{COS}} & \textit{\textbf{WMD}} & \textit{\textbf{COS}} & \textit{\textbf{WMD}} & \textit{\textbf{COS}} & \textit{\textbf{WMD}} & \textit{\textbf{SCM}} & \textit{\textbf{COS}} & \textit{\textbf{EUC}} \\ \hline
 & \stlibest & 0.44 & 0.28 & -0.16 & -0.13 & 0.12 & 0.14 & 0.69 & 0.33 & 0.41 & -0.19 \\
& \stcisco & \cellcolor[HTML]{DAE8FC}0.74 & 0.01 & -0.63 & -0.12 & 0.49 & 0.05 & 0.63 & 0.5 & 0.22 & -0.37 \\
\multirow{-3}{*}{\exbase} & \stdronology & 0.29 & -0.09 & -0.15 & -0.12 & -0.08 & 0.08 & 0.26 & 0.23 & 0.06 & -0.23 \\\hline
 & \stlibest & 0.62 & 0.02 & -0.52 & -0.03 & 0.59 & 0.01 & \cellcolor[HTML]{DAE8FC}0.71 & 0.67 & 0.01 & 0.05 \\
& \stcisco & \cellcolor[HTML]{DAE8FC}0.87 & 0 & \cellcolor[HTML]{DAE8FC}-0.81 & 0 & 0.58 & 0 & \cellcolor[HTML]{DAE8FC}0.87 & \cellcolor[HTML]{DAE8FC}0.81 & -0.01 & -0.03 \\ 
\multirow{-3}{*}{\exfirst} & \stdronology & 0.38 & 0.09 & -0.3 & -0.27 & -0.05 & 0.08 & 0.43 & 0.39 & 0.27 & -0.34\\ \hline
 & \stlibest & 0.66 & 0.02 & -0.53 & 0.01 & \cellcolor[HTML]{FFCCC9}0.62 & 0 & \cellcolor[HTML]{DAE8FC}0.75 & 0.52 & 0.03 & -0.05 \\
& \stcisco & \cellcolor[HTML]{DAE8FC}0.86 & 0 & \cellcolor[HTML]{DAE8FC}-0.79 & 0 & \cellcolor[HTML]{FFCCC9}0.62 & 0 & \cellcolor[HTML]{DAE8FC}0.83 & 0.76 & 0 & -0.03 \\ 
\multirow{-3}{*}{\exsecond} & \stdronology & 0.38 & 0.25 & -0.28 & -0.24 & -0.05 & 0 & 0.41 & 0.34 & 0.27 & -0.28\\ \hline
 & \stlibest & 0.47 & 0.15 & -0.2 & -0.1 & 0.17 & 0.06 & \cellcolor[HTML]{DAE8FC}0.71 & 0.4 & 0.46 & -0.12 \\
 & \stcisco & \cellcolor[HTML]{DAE8FC}0.78 & 0.34 & -0.67 & -0.27 & 0.53 & 0.18 & 0.65 & 0.5 & 0.42 & -0.31 \\ 
\multirow{-3}{*}{\exthird} & \stdronology & 0.29 & 0.17 & -0.1 & -0.03 & -0.16 & -0.13 & 0.24 & 0.23 & 0.27 & -0.27\\ \hline
 & \stalbergate & 0.25 & \cellcolor[HTML]{DAE8FC}0.35 & \cellcolor[HTML]{FFCCC9}0.43 & \cellcolor[HTML]{FFCCC9}0.41 & -0.2 & -0.37 & 0.56 & 0.45 & 0.59 & \cellcolor[HTML]{FFCCC9}-0.72 \\
 & \stebt & 0.36 & -0.04 & -0.04 & -0.19 & -0.14 & 0.05 & 0.26 & 0.22 & 0.21 & -0.29 \\
 & \stetour & 0.39 & 0.25 & 0.07 & 0.05 & \cellcolor[HTML]{DAE8FC}-0.23 & -0.13 & 0.59 & 0.45 & 0.44 & -0.56 \\
 & \stitrust & 0.54 & 0.13 & -0.32 & -0.12 & 0.13 & -0.01 & 0.61 & 0.44 & 0.3 & -0.44 \\
\multirow{-5}{*}{\exfourth} & \stsmos & \cellcolor[HTML]{FFCCC9}0.19 & \cellcolor[HTML]{FFCCC9}-0.09 & -0.39 & 0.14 & 0.31 & 0.2 & 0.4 & 0.48 & 0.2 & -0.28 \\ \hline
\end{tabular}
}

{
\vspace{0.2em}
\scriptsize\textit{Red cells indicate the lowest value of each metric; blue cells indicate the highest.}
}
\end{table}

%%%%% GRID IMAGES

\textit{Mutual Information \& Shared Information Entropy and Extropy.} This analysis correlates the distance and entropy measures. As shown in \tabref{tab:correlations}, mutual information is negatively correlated with \wmd distance (equivalently, positively correlated with \wmd similarity): more shared information implies smaller artifact distance. MI shows no comparable correlation with cosine similarity; because \wmd is reported alongside \wv and \cosine alongside \dv in our experimental design (\tabref{tab:performance}), this contrast reflects the combined effect of the distance metric and the embedding technique, and we cannot attribute it to either factor in isolation, suggesting that word vectors may capture semantic relationships better than paragraph vectors; though both ultimately underperform on the binary classification. The MSI for entropy is also negatively correlated with \wmd (\tabref{tab:correlations}), mirroring the trend observed for mutual information and reinforcing  evidence that \wmd similarity better captures semantic relationships among artifacts.

\textit{Composable Manifolds.} The composable manifolds allow inspection of a third information variable; we focus here on loss and noise (\figref{fig:grid2}). On \cisco, loss is largest where MI and similarity are lowest (\figref{fig:sub53}), while noise is more dispersed and forms clusters at both low and high MI (\figref{fig:sub54}), suggesting that injected information is independent of artifact semantics. On \libest (\figref{fig:sub51}, \figref{fig:sub52}), the \wmd–MI correlation is dispersed: loss stays high at low MI, noise at high MI. On \dronology (\figref{fig:sub55}, \figref{fig:sub56}), MI entropy is more condensed, entropy loss is dispersed, and noise concentrates at low MI.

\vspace{-0.5em}
\begin{boxK}
%\vspace{-0.5em}
    \ref{rq:correlation} Loss entropy correlates with low similarity and mutual information, suggesting a clear intervention point: reducing loss in traceability datasets improves link classification. Such refactorings are feasible in practice, as practitioners routinely complete and document artifacts across software life cycle.
%\vspace{-1.5em}
\end{boxK}
\vspace{-0.2em}

\section{A Case Study in Industry}\label{sec:cases}

%This section shows four information science cases from processing the \cisco testbed. Experiments and samples can be found in our online appendix~\cite{palacio2023tracexplainer,RepoTraceXplainer24}.

This section presents four information-science cases derived from processing the \cisco testbed, an industrial corpus of pull requests and source code drawn from a real-world Cisco engineering workflow.  Each case is framed around an actionable insight that engineering teams can apply to assess and improve their traceability readiness. The complete experimental artifacts and supporting samples are available in our online appendix~\cite{palacio2023tracexplainer,RepoTraceXplainer24}.

\textit{$Case_0$: Diagnosing Information Imbalance Between Pull Requests and Code.} This case surfaces a recurring industrial pain point: the information asymmetry between source artifacts (\eg pull requests, issue descriptions) and target artifacts (\eg production source code). In the \cisco workflow, pull request comments exhibit an average self-information of $3.42\,\mathcal{B}$, while the corresponding source files reach $5.91\,\mathcal{B}$, indicating that code carries substantially more information than the pull request intended to describe it. Artifacts with low entropy; typical of terse PR titles, single-line tickets, or boilerplate templates, fail to provide unsupervised techniques with the lexical density needed to recover trace links. Conversely, artifacts with disproportionately high entropy introduce loss and noise that obscure the underlying engineering intent. In practice, this imbalance is a leading indicator that documentation conventions, PR templates, or commit hygiene practices require attention before model-centric investments will yield returns. A related industrial consideration is that source code is consumed by these unsupervised pipelines as plain text rather than as a structured language; the resulting information loss from tokenizing code as a sequence has not yet been quantified and represents a concrete improvement opportunity for applied research teams.

%\textit{$Case_0$: Self-Information.} This study highlights the information imbalance between the source and target artifacts. Artifacts with low entropy struggle to generate traceability links since unsupervised techniques rely on concise descriptions in natural language. By contrast, high levels of entropy struggle with other conditions like loss or noise. Refactoring operations need to be applied to source and target artifacts to combat the information imbalance. At least, a semantic idea expressed as a clause is required in both artifacts to guarantee a link. When we preprocess the source code, we transform the code structure into a sequence structure to extract those clauses. In other words, source code is not treated as a structured language but as a regular text file. The amount of information lost by preprocessing source code as text has not yet been computed.

\textit{$Case_1$: Detecting Under-Documented Code via Loss Extremes.} This case examines the edge cases of entropy loss to identify production code that is not adequately described in the originating pull request. Because loss, as assumed by default in \secref{sec:approach}, conflates genuine documentation gaps with the ordinary lexical divergence between natural-language and code vocabularies, we treat the extremes as candidates for manual triage rather than confirmed documentation defects. For practitioners, these extremes are the highest-leverage refactoring targets in the target documentation. At the upper extreme, a pull request consisting of a single token (\eg a one-word commit message) is paired with one of the highest-entropy source files in the repository, a clear signal that the change description fails to capture the implementation. At the lower extreme, a fully described pull request is paired with source code whose vocabulary does not reflect the PR content; even a human reviewer attempting manual link recovery (at the $99\%$ quartile for positive links) would struggle to justify the association. In an industrial setting, both extremes warrant intervention: the first via stricter PR description standards, the second via inline code comments or naming conventions that mirror the requirement vocabulary.

%\textit{$Case_1$: Minimum and maximum loss.} This study presents edge cases for entropy loss to identify poorly documented target artifacts. Edge cases of loss are useful to detect starting points for general refactoring in the target documentation. The max case shows us the pull request content composed of just one word. The tokens that represent this word were not found in the target artifact, which is one of the highest entropy files. Whereas, the min case shows us the pull request has a complete description not found in the target. Note the 99\% quartile for positive links. The PR content cannot be easily found in the source code with low entropy, even if a practitioner tries to extract links manually. %The relationship is not evident.

\textit{$Case_2$: Detecting Under-Documented Requirements via Noise Extremes.} This case analyzes entropy-noise edge cases to identify source artifacts (pull requests, requirements, or change tickets) that fail to ground themselves in the corresponding code. In the maximum-noise scenario, a content-rich pull request is associated with an essentially empty target file, suggesting that the implementation is trivial, generated, or that the link itself is spurious. In the minimum-noise scenario, the pull request content is repetitive and lexically narrow, providing little discriminative signal for any downstream automation. For engineering leadership, these patterns highlight two distinct documentation debts: empty targets that require code-level commenting and repetitive PR narratives that point to template fatigue or copy-paste workflows in the change-management process.

%\textit{$Case_2$: Minimum and maximum noise.} This study presents edge cases for the entropy noise to identify poorly documented source artifacts. Edge cases of noise are useful to detect refactorings in documentation found in the source documentation. In the max case, the PR content is high, but the target artifact is empty. This suggests that the target is not well documented. Something similar occurs in the min case, where the PR content is repetitive and less expressive.

\textit{$Case_3$: Surfacing Orphan Informative Links Missing From Ground Truth.} This case identifies link candidates that exhibit strong information-theoretic alignment between source and target artifacts but are absent from the recorded ground truth. From an industrial quality-assurance perspective, orphan links are doubly valuable: they expose inconsistencies and gaps in the trace matrix maintained by the engineering team; which may have implications for compliance, audit, and impact analysis, and they surface likely-true links that hold independently of the unsupervised technique applied. Embedding this analysis into a continuous integration pipeline would give release managers and compliance owners an early-warning signal for incomplete traceability records, well before downstream activities, such as security tracking or change-impact reviews, are affected.

%\textit{$Case_3$: Orphan informative links.}  This study points out a set of informative links not found in the ground truth. Orphan links not only exhibit inconsistencies in the ground truth file but also suggest potential positive links independent of the employed unsupervised technique.

%\input{texts/7.relatedWork}

\section{Threats to Validity}\label{sec:threats}

\textbf{Internal Validity.} Trace-link ground truth for all eight testbeds was provided by project maintainers; our own analysis suggests some labels are imperfect, \eg the noise measure (\secref{sec:approach}) flags likely mislinked configuration files, and $Case_3$ (\secref{sec:cases}) surfaces high-MI candidate links missing from the ground truth. \ref{rq:effectivess} and \ref{rq:semantic} treat these labels as reference, reported effectiveness should be read as approximate rather than exact.

\textbf{Construct Validity.}\approach characterizes the informativeness and lexical alignment of artifacts but does not predict or guarantee the correctness of individual trace links; its measures should be interpreted as risk indicators. This follows from its copy-style transmission assumption (\secref{sec:approach}), whereas most links involve transformations between natural-language and programming-language representations with limited vocabulary overlap. As mutual information cannot distinguish these regimes, it may underestimate valid but lexically divergent links. Accordingly, loss and noise values should be treated as relative signals, not absolute measures of quality. We therefore restrict our claims to \textbf{lexical (token-level) traceability} and position \approach as a complementary diagnostic tool.

\textbf{External Validity.} We evaluate \approach only against classical unsupervised embedding techniques (\wv, \dv) rather than stronger modern paradigms, \eg supervised rankers, code-language-model rerankers, graph-based methods, or LLM/RAG-based approaches (\secref{sec:background}). Whether these paradigms can partially overcome the data-centric ceiling we report is an open empirical question. The four cases in \secref{sec:cases}  are not general-purpose diagnostic procedure, and their generalization to other systems or domains remains untested.

\section{Lessons Learned for Industry}\label{sec:lessons}

Our empirical study and industry-oriented case analysis yield three practical lessons for deploying unsupervised traceability in real-world settings, supporting a shift from model-centric optimization toward data-centric practices that prioritize artifact quality, consistency, and informational richness.
\textbf{First, traceability performance is primarily constrained by artifact informativeness and alignment, not model sophistication.} Across all configurations, \wv and \dv fail when artifacts lack sufficient or overlapping information. In practice, improving requirements, pull requests, and documentation through clearer structure, consistent terminology, and completeness offers greater gains than further model tuning.

\textbf{Second, information-theoretic discrepancies (\ie loss and noise) provide actionable signals of misalignment.} High loss indicates missing propagation of source information, while noise reflects undocumented or extraneous target behavior. These signals correlate with weak traceability and can guide refactoring, documentation improvements, and quality assurance efforts.

\textbf{Third, standard evaluation metrics alone are insufficient.} Measures such as precision, recall, and AUC can obscure limitations due to imbalance or low-information artifacts. Information-theoretic metrics (\eg entropy and mutual information) offer complementary insight into whether traceability is feasible and help diagnose failure modes.

%Overall, these findings support a shift from model-centric optimization to data-centric practices that prioritize artifact quality, consistency, and informational richness.

 %\revision{In our \cisco analysis, the average level of mutual information was $3.21\,\mathcal{B}$ (error $0.02$ at a $95\%$ confidence level); we report this figure as a descriptive statistic for this testbed and experimental configuration, not as a threshold or rule of thumb for other systems. Because mutual information depends on vocabulary size, artifact length, and domain (\secref{sec:design}), we do not recommend adopting $3.21\,\mathcal{B}$, or any other fixed value reported in this paper, as a general traceability-risk threshold. Instead, we recommend that engineering teams establish a system-specific baseline, \eg the median MI across their own confirmed links, and treat substantial, relative deviations from that baseline as the actionable signal, consistent with our broader guidance that \approach's measures are comparative rather than absolute (\secref{sec:lessons}).} 

\section{Acknowledgments}
\label{sec:acknowledgments}

%This research has been supported in part by XXXX. We also acknowledge support from YYYY. Any opinions, findings, and conclusions expressed herein are the authors’ and do not necessarily reflect those of the sponsors. 
This research has been supported in part by NSF CCF-2346357, CCF-231146 and CCF-2423813 grants. We also acknowledge support from Cisco Systems. %Any opinions, findings, and conclusions expressed herein are the authors’ and do not necessarily reflect those of the sponsors. 

%%%%%%%%%%%%%%%%%%%%%%%%%%% REFERENCES
\bibliographystyle{abbrv}
\bibliography{utils/info_theory}

\end{document}